# A Robust and Fault-Tolerant Distributed Intrusion Detection System

Jaydip Sen
*Innovation Lab, Tata Consultancy Services Ltd., Kolkata, India*
*Jaydip.Sen@tcs.com*

**Abstract**

*Since it is impossible to predict and identify all the vulnerabilities of a network, and penetration into a system by malicious intruders cannot always be prevented, intrusion detection systems (IDSs) are essential entities for ensuring the security of a networked system. To be effective in carrying out their functions, the IDSs need to be accurate, adaptive, and extensible. Given these stringent requirements and the high level of vulnerabilities of the current days' networks, the design of an IDS has become a very challenging task. Although, an extensive research has been done on intrusion detection in a distributed environment, distributed IDSs suffer from a number of drawbacks e.g., high rates of false positives, low detection efficiency etc. In this paper, the design of a distributed IDS is proposed that consists of a group of autonomous and cooperating agents. In addition to its ability to detect attacks, the system is capable of identifying and isolating compromised nodes in the network thereby introducing fault-tolerance in its operations. The experiments conducted on the system have shown that it has high detection efficiency and low false positives compared to some of the currently existing systems.*

## 1. Introduction

There have been two different approaches for securing networks and host computers from malicious attackers: i) intrusion prevention mechanisms that include cryptographic techniques to safeguard sensitive information from unauthorized access, ii) intrusion detection mechanisms that recognize an ongoing attack on a system and respond appropriately to thwart such intrusive attempts. An *intrusion detection system* (IDS) is a security mechanism that can monitor and detect intrusions to the computer systems in real time. An IDS can be either host-based or network-based or a combination of both. The conventional approach to intrusion detection involving a central unit to monitor an entire system has several disadvantages [1]. To circumvent the demerits of a centralized IDS, the research in the field of intrusion detection over the last decade has been heading towards a distributed framework. In these systems, the local intrusion detection components look for local intrusions and pass the results of their analyses to the upper levels in the hierarchy. The components at the upper levels analyze the refined data from multiple lower level components and seek to establish a global view of the system state.

In this paper, the scheme of a distributed IDS is presented. This is an extension of our earlier work presented in [2]. In the proposed IDS, the knowledge about attack scenarios is introduced into a large number of agents in the form of a Bayesian network so that each agent is required to monitor only a relatively few aspects of the local network. However, the agents share their beliefs, and through timely coordination of the agents, the system is able to detect complex distributed attacks. For distributed inference, the concept of *multiply sectioned Bayesian networks* (MSBN) [3] is used in domain knowledge representation. To reduce network congestion and message overhead, the agents are grouped into subdomains. The communications among the agents are mostly confined in the subdomains.

The main contributions of the paper are as follows. It presents a scheme for intrusion detection that provides: (i) efficient detection of intrusive activities by local monitoring and sharing of individual belief estimates of the agents for a collaborative detection, (ii) fault-tolerance by prompt identification and isolation of compromised hosts in the network through a distributed trust framework, and (iii) better detection efficiency and reduced false positives while providing enhanced robustness and fault-tolerance compared with the existing intrusion detection schemes.

The organization of the paper is as follows. Section 2 describes the architecture of the system, agents and their communication framework. Section 3 discusses Bayesian intrusion detection framework used in the proposed system. Section 4 presents a distributed trust framework that enhances the fault-tolerance of the security mechanism. Section 5 presents the experimental results. Section 6 concludes the paper while identifying some future scope of work.

## 2. Architecture of the system and agents

In this section, the architecture of the system and its constituent agents are described. The proposed system consists of a large number of agents that collect data, analyze, and make appropriate inference. The agents use an inference process that utilizes a Bayesian network of data.

The agents are grouped into several subdomains. The agents in the same subdomain communicate actively and frequently. Communication between agents belonging to adjacent subdomains happens infrequently. The agents have knowledge about a Bayesian network model of the structures of well-known attack types as well as normal usage pattern, which is constructed offline from data repositories containing system logs from ongoing attacks. The global Bayesian network is partitioned into multiple subnets based on the spatial locations of the agents.

Bayesian network approach allows for combining anomaly detection and signature recognition. For this purpose, one Bayesian network is generated whose nodes classify known attack types and normal behavior. Using this network, the agents can detect normal behavior and known attacks. If the probabilities associated with none of the target nodes cross the threshold, the system detects an anomalous behavior.

Every host in the system has one special agent, called the *distributed trust manager* (DTM). It uses the Byzantine Agreement Protocol among the peers and identifies any compromised host in the system. This distributed trust mechanism makes the proposed system robust and fault-tolerant.

### 2.1. Agent architecture

Figure 1 depicts the architecture of an agent. Each agent consists of six modules. The functionalities of each of these modules are briefly described below.

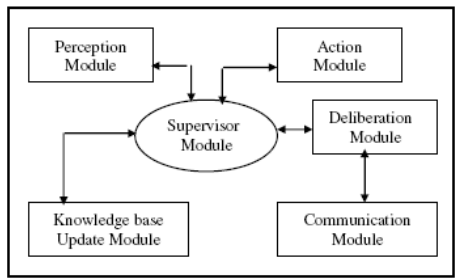

**Figure 1: Architecture of an agent in a node**

*Perception module* is responsible for collection of audit or network data of the subdomain to which the agent belongs. *Deliberation module* is responsible for analyzing the data collected by the perception module. It enables an agent to reason and extrapolate by relying on built-in knowledge and allows the agent to update its belief in the subnet it is monitoring. *Communication module* allows an agent to communicate its belief, decisions, and knowledge to its peer agents in the same subdomain, and possibly, to other subdomains also. *Action module* takes appropriate actions when an intrusion is detected. When an agent recognizes that a monitored host has exceeded the threshold of one known attack, it triggers an alert and communicates it to the system administrator. The agents can also trigger an alert indicating an *anomalous* situation, when the activated target node does not belong to those representing normal behavior or any of the known attack types. The system administrator can either confirm the attack (or take necessary steps to handle it), or reject the alert if it is found to be a false alarm. *Knowledgebase update module* updates the attack signature database if the system administrator confirms an anomaly alert. The Bayesian network is modified to accommodate this new attack in the knowledge base. *Supervisory module* is the central module that coordinates the tasks and interactions among the other modules.

### 2.2. Communication system of agents

As shown in Figure 2, three types of agents are deployed in the proposed system. The *system monitoring agents* collect, transform, and distribute intrusion specific data when requested and evoke information collection procedures. The *intrusion monitoring agents* subscribe to beliefs published by the system monitoring agents and other intrusion monitoring agents. Each intrusion monitoring agent has the knowledge of a local Bayesian network of attacks. These agents update their knowledge on receipt of information from other agents. For each registered agent, a registry is maintained for the variables it monitors. The agents use the registry to find the names and locations of agents, which may have data that they are interested in. The messages are in XML format and may be of different types e.g. : (i) registration of agents with registry agents, (ii) request to registry agents for locations of other agents, (iii) search of agent queries, iv) belief subscription requests, v) belief updates.

Figure 3 shows two different types of communication among the agents. Various mechanisms exist for these communications [4]. In the proposed mechanism, the shared memory architecture is used for agent communication since it allows large volume of data to be shared among agents. For efficient communication among agents over the network, an *agent management system* (AMS) is used. The capabilities provided by the *java agent development environment* (JADE) [4] framework has been utilized to build an AMS.

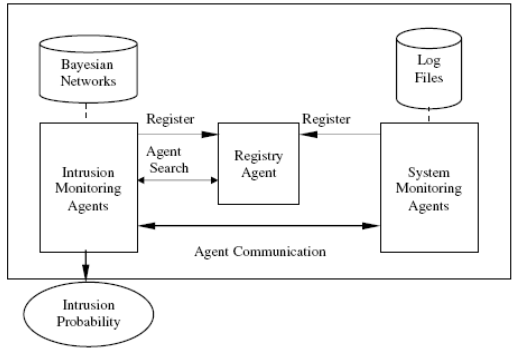

**Figure 2: Major components of the system**

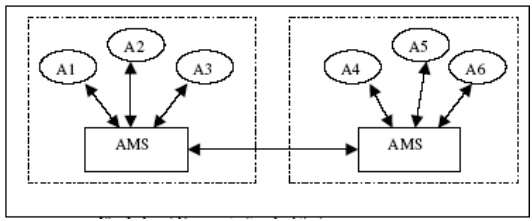

**Figure 3: Agent communication architecture**

For security, *public key infrastructure* (PKI) is used to provide two-way authentication of the agents and the messages. The messages are encrypted using symmetric key cryptography.

## 3. Bayesian Intrusion Detection

This section describes how the concepts of Bayesian networks and MSBNs are applied in the proposed IDS. As mentioned in Section 3, the knowledgebase of the attacks are distributed in the agents in the form of a Bayesian network. A Bayesian network is a *directed acyclic graph* (DAG) in which the nodes represent the variables, and each directed edges represents a dependency between the corresponding variables. The effect of the parents of a node on a node is represented by conditional probabilities of that variable given values of its parent nodes in the form of a *conditional probability table* (CPT). There are several reasons for using Bayesian network in the proposed system. First, a Bayesian network can handle incomplete information, which is suitable for the agents that have limited local view of the network and may receive only partial information about a possible attack. Secondly, a Bayesian network can represent causal relationships among variables, which can help an intrusion detection model to combine a priori knowledge and observed data to take a decision. Finally, a Bayesian network allows updating of the beliefs and thus can be used to recognize novel attack signatures. In the proposed system, a Bayesian network is first constructed from a database of known attacks, which is then partitioned into several subtrees following the principle of MSBNs [3], and distributed among the agents. This process is explained in the following subsection.

### 3.1. Inference with MSBNs

An MSBN consists of a set of interrelated Bayesian subnets each of which encodes an agent's knowledge of a subdomain. In such a framework, probabilistic inference can be performed in a distributed fashion. Multiagent inferences in MSBNs are usually done using message passing in *junction trees* (JTs) [3]. The *linked junction forest* (LJF) method compiles each subnet of a multiply connected network into a JT by clustering the triangulated moral graph of the underlying undirected graph. The algorithm performs message propagation over the JT. Message passing among agents in two adjacent subnets is performed through a *linkage tree*. Though belief update in a Bayesian network is NP-hard [5], it can still be used for intrusion detection since the subnets are usually small.

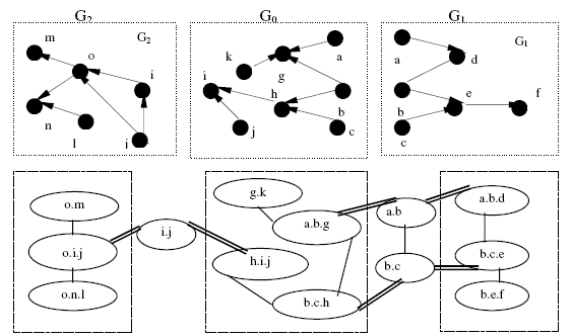

**Figure 4: DAGs of three subnets of an MSBN and JTs constructed from the subnet**

Figure 4 shows an MSBN with three subnets $G_0$, $G_1$, $G_2$. Each of these subnets contains a group of agents. The LJF method has compiled each subnet into a JT (called a local JT), and has converted each d-sepset into a JT (called a linkage tree). Figure 4 also depicts three local JTs and two linkage trees of the monitoring system. Each oval in a JT or a linkage tree represents a subset of variables and is called a *cluster*. For instance, $\{o, i, j\}$ is a cluster in a JT in the subnet $G_2$ and $\{i, j\}$ is a cluster in the linkage tree between subnets $G_2$ and $G_0$. Local inference is performed by message passing in the local JT. Message passing between a pair of adjacent subdomains is performed using the linkage tree.

Once a multiagent MSBN is constructed, agents perform probabilistic inference by computing the query $P(x|e)$, where $x$ is any variable within a subdomain and $e$ is the observations made by all the agents in the system. The system-wide communication among the agents (about $e$) is vital for sharing of information among agents. However, since the agents are autonomous, the system-wide message passing is infrequent. Most of the time, the agents in subnet $G_i$ computes the query $P(x|e_i, e_i')$, where $e_i$ is the local observations made by the agents in $G_i$, and $e_i'$ is the observations made by the agents of other subdomains as recorded in $G_i$ till the last communication. This computation is called *local inference*. Since LJF has least overhead of inter-agent communication [6], in the proposed system it has been used to minimize network traffic due to inter subdomain message passing.

## 4. Fault-Tolerance and Trust Mechanism

In this section, a novel approach for introducing fault-tolerance in the proposed system is described. A distributed trust management scheme is developed among the agents in the system and a robust algorithm based on *Byzantine Agreement Protocol* (BAP) [7] is invoked among the peer agents. This enables a reliable and fast detection of any compromised agents in the system. If any agent is detected to be compromised, it is immediately isolated from the system. Use of BAP to identify compromised nodes makes the proposed IDS framework reliable, secure and fault-tolerant.

### 4.1. Distributed trust management

The agents are always vulnerable to attacks by intruders. If an intruder can compromise a host, the agents in the host will attempt to influence the JT and which will adversely affect the overall inference machinery of the IDS. For early detection of compromised hosts in the proposed system, an efficient trust management scheme based on BAP is developed among the hosts.

### 4.2. The Byzantine agreement protocol

Lamport et al. described the Byzantine Generals Problem in [7]. The problem is as follows: Imagine that several divisions of a Byzantine army are camped outside an enemy city, each division commanded by its own general. The generals can communicate with each other only by messengers. After observing the enemy, they must arrive at a common plan of action. However, some of the generals may be traitors, and try to prevent the loyal generals from reaching an agreement. The generals must have an algorithm to guarantee that all loyal generals decide upon the same plan of action. The loyal generals will all do what the algorithm asks them to do; however, the traitors may do anything they wish.

The BAP is essentially an algorithm designed to achieve consensus among a set of processes in a distributed system. These processes can arrive at a consensus if they all agree on some allowed values called the *outcome* (if they could agree on any value the solution would be trivial: always agree on '0'). Thus arriving at a consensus involves two actions: first specify a value, and the read the outcome of execution of the processes involved. The consensus algorithm terminates when all non-faulty processes come to know the outcome. If we consider the generals in BAP as the hosts in a distributed system, and the consensus as the requirement of agreement among the hosts as which hosts are safe (i.e. not compromised), then the problem of identifying compromised host(s) in a distributed can be described more formally as follows. Consider a distributed system consisting of a large number of hosts, with each host having a group of agent. The agents cooperate to detect malicious intrusive attempts or actual intrusions into the system. Each host has a special agent - *distributed trust manager* (DTM), which continuously sends messages to its peers. The messages sent by the DTM are of two types: (i) *Message $A_1$:* the host is safe with a value '0', (ii) *Message $A_2$:* the host is compromised with a value, '1'.

Lamport et al. proposed a *signed message algorithm* (SMA) for solution of the Byzantine Generals Problem which requires $O(n^2)$ messages for $n$ hosts. It works effectively if there are at most $n$ - 2 traitors (i.e. compromised hosts). However for SMA to work correctly, some conditions need to be satisfied: (i) each message is delivered correctly, (ii) the receiver of a message knows its sender, (iii) the absence of a message in the buffer of a host can be detected, (iv) the signature of a safe (i.e. not compromised) host cannot be forged, and (v) any host can verify the authenticity of its general's signature.

With cryptographic mechanisms in place, the proposed system guarantees that all the above conditions are satisfied and hence SMA will work well.

In the SMA, one of the hosts acts as the leader and sends an order to the other hosts. Whenever a host receives a message, it takes the order and puts it in its list of the orders received. Then the receiver signs the message with its own signature and forwards it to all the hosts whose signature is not on the order. If a host receives a message with an order that is already in his list, it ignores the message. When no more messages are left to be received, all the hosts choose an order from the list of orders they have received using this method. If only one order has been received, that order is chosen. Because any order that reaches a safe host will be forwarded to all other hosts who have not seen the order, all the safe hosts will have the same set of orders to choose from, and thus choose the same order to obey.

### 4.3. Distributed trust manager

DTM forms and maintains *trust domains*. A trust domain is a set of hosts that share a charter and a security policy and behave consistently in accordance with that policy. The hosts in a trust domain ensure that no compromised host can join the trust domain. To understand the functions of DTM, we assume that at the beginning of a trust domain formation, all the hosts are safe. Any compromised host in the trust domain is identified by running $n$ instances ($n$ is the number of hosts in the trust domain) of SMA in parallel, assuming that the majority of the hosts in the trust domain are not compromised. If the *leader* of the SMA is not compromised, then after running the algorithm in parallel, all the hosts that are not compromised will know that the leader is not compromised. However, if the leader happens to be compromised, then one of the following possible cases may happen: (i) The leader sends '*0*' to all the safe hosts. In this case, all the safe hosts will assume the leader host to be compromised or dead. (ii) The leader sends '*1*' to only some of the safe hosts. In this case, those safe hosts that have received '*1*' from the leader are able to detect that there is a compromised host in the system. These hosts, then, send messages to other hosts informing about the suspected compromised host. On further investigation based on additional message communication, the actual status (compromised or otherwise) of the suspected node would be understood. (iii) The leader sends '*1*' to all safe hosts. The safe hosts can now realize that the message is wrong, and the leader itself is compromised if it contradicts the majority. If the message does not contradict the majority, it is not possible to conclude about the status of the leader unless it sends a different message to at least one compromised host, which in turn forwards the message to a safe host. In the latter case, the leader host is compromised and is isolated. (iv) The leader sends two (or more) different messages to some of the safe hosts. The all the safe hosts together find contradictory instructions, and understand that the leader is compromised. In this way, DTM can detect compromised nodes in the system in all possible cases.

## 5. Experiments and Results

A proof-of-concept prototype for the proposed IDS has been built using Java and the JADE [4] environment. JADE is a middleware for enabling faster development of multi-agent distributed applications based on the peer-to-peer communication architecture.

In the prototype IDS, each agent is endowed with three behavioral capabilities: *filtering*, *interaction*, and *deliberation*. The filtering behavior of an agent enables it to filter security events from the observations it makes. The interaction behavior manages the interaction of the agent with its peers and defines the way the messages are received and enqueued. The deliberation behavior allows an agent to represent its beliefs, goals, intentions, and knowledge in a semantic format. When an agent receives a detection goal, it updates a set of event classes to filter. When an event occurs, it is filtered by the filtering module and sent to the deliberation module. The deliberation module updates the agent's beliefs, and checks whether the belief matches with an attack signature. If it matches, a detection goal is reached and a list of intentions is sent to the interaction module for execution.

Experiments have been conducted to test the performance of the proposed IDS. The KDD Cup 1999 intrusion detection contest data [8] has been used in the experiments. The original data contains 744 MB of information with 4.94 million records. The dataset has 41 attributes for each connection record plus one connection record specifying one of 24 different types of attacks or normal condition. Thus, effectively each record is given a class label that specifies the category of attack to which the record belongs. All the attacks are grouped into 4 major categories: (i) *denial of service* (DoS), (ii) *remote to local* (R2L), (iii) *user to root* (U2R), and (iv) *probe*.

A dataset of 15000 records is constructed by randomly selecting records from the original database, such that the number of data instances selected from each class was proportional to their frequencies in the original database. An additional class called the *normal* class is also constructed. The attack knowledge base is distributed among the agents, and a *Bayesian network power constructor* (BNPC) [9] is used to generate a Bayesian network from these sampled records. This Bayesian network is sectioned into multiple subnets utilizing the rules for sound partitioning in MSBN [3]. Finally, the LJF method is used for intrusion detection.

The prototype is tested in a network of 50 workstations. Each workstation has Pentium 4 processor, 3GHz clock speed, 1 GB RAM, and Linux operating system. The interconnecting medium is Ethernet with 100 mbps capacity. Using the *Ethereal* network sniffer, memory, CPU cycle and bandwidth consumption due to the agents are evaluated. From Table 1, it is evident that the average memory and the CPU usage on a workstation due to the IDS are marginal and the overhead decreases with increase in number of users. The maximum CPU usage by the agents is found to be 8.76% with the average being 5.34%. During the period when the agents were active, only 15% increase in number of packets is found in the network. The average bandwidth consumed by the agents never exceeded 5%.

**Table 1. Impact of the IDS on user applications**

| No. of users | 10 | 20 | 30 | 40 | 50 |
|---|---|---|---|---|---|
| Mem. reqd with IDS (K) | 331 | 471 | 601 | 710 | 795 |
| Mem. reqd. without IDS (K) | 324 | 453 | 572 | 669 | 740 |
| CPU usage with IDS (%) | 53 | 67 | 80 | 86 | 92 |
| CPU usage without IDS (%) | 48 | 63 | 75 | 82 | 87 |

For testing the detection efficiency of the proposed IDS and its false positives rates, 37 different attacks are simulated in the network on different workstations. Some of attacks are chosen in such a way that they are not in the knowledge base of the agents. This is done to test the ability of the IDS prototype to detect novel attacks. The detection efficiency of the proposed IDS is also compared with the scheme proposed by Li et al in [10]. From Table 2, it is evident that except for the 'normal' scenario, the detection rate of the proposed IDS is better than that of Li et al.'s scheme [10].

**Table 2. Operational performance of the IDS**

| Activity | Detection Rate (%) | | False Positive (%) |
|---|---|---|---|
| | Proposed IDS | Li et al's IDS | Proposed IDS |
| DoS | 98.25 | 97.57 | 10.25 |
| R2L | 7.31 | 0.37 | 12.43 |
| U2R | 86.42 | 71.49 | 10.57 |
| Probe | 94.28 | 90.49 | 11.87 |
| Normal | 97.80 | 98.13 | 7.31 |

The better performance of the proposed scheme is due to the robust knowledge base building and inference mechanism of the JADE framework. However, like most of the existing IDS schemes, the proposed scheme has a low detection rate for R2L attacks. Although, the detection rate for these attacks is much higher compared to Li et al.'s scheme, it is far from satisfactory. Since R2L attacks are essentially different from the other types of attacks, there is a need for a different approach for detection logic. The DTM scheme is mostly responsible for the higher detection rates in DoS and probe attacks. The false positive rates are also found to be fairly low for all categories of attacks.

# 6. Conclusion

This paper has presented the framework of a distributed IDS that consists of a group of agents cooperating with each other to carry out the task of intrusion detection. Using distributed computation and message passing, the agents detect both signature-based attacks and anomalous activities in real-time. Apart from its ability to make distributed inference, the proposed IDS can also identify compromised nodes in the system with the help of Byzantine Agreement Protocol. The experiments show that the performance of the IDS is better than some of the currently existing systems. Development of a new detection logic for R2L type of attacks constitutes a future plan of work.